\documentclass[runningheads]{llncs}
\usepackage{graphicx}
\usepackage{multirow}
\usepackage{wrapfig}
\usepackage{booktabs}
\usepackage{multirow}
\usepackage{caption}
\usepackage{float}
\usepackage{bm}
\usepackage{nicefrac}
\usepackage{hyperref}
\usepackage{lipsum}
\usepackage{multirow}
\newcommand\blfootnote[1]{%
  \begingroup
  \renewcommand\thefootnote{}\footnote{#1}%
  \addtocounter{footnote}{-1}%
  \endgroup
}

\begin{document}
\title{Allegro-Legato: Scalable, Fast, and Robust Neural-Network Quantum Molecular Dynamics via Sharpness-Aware Minimization}

\titlerunning{Allegro-Legato}
%

\author{Hikaru Ibayashi \inst{1}\and
  Taufeq Mohammed Razakh \inst{1}\and
  Liqiu Yang \inst{1}\and
  Thomas Linker \inst{1}\and
  Marco Olguin \inst{2} \and
  Shinnosuke Hattori \inst{3}\and
  Ye Luo \inst{4}\and
  Rajiv K. Kalia \inst{1}\and
  Aiichiro Nakano \inst{1}\and
  Ken-ichi Nomura \inst{1}\and
  Priya Vashishta \inst{1}}
\authorrunning{H. Ibayashi et al.}
\institute{
  Collaboratory for Advanced Computing and Simulations, University of Southern California, Los Angeles, CA 90089, USA
  \and
  Center for Advanced Research Computing, University of Southern California, Los Angeles, CA 90089, USA
  \and
  Advanced Research Laboratory, R\&D Center, Sony Group Corporation, Atsugi Tec. 4-14-1 Asahi-cho, Atsugi-shi, Kanagawa 243-0014, Japan
  \and
  Argonne Leadership Computing Facility, Argonne National Laboratory, Lemont, IL 60439, USA
}
\maketitle              
\vspace{-0.73cm}
\begin{abstract}
  Neural-network quantum molecular dynamics (NNQMD) simulations based on machine learning are revolutionizing atomistic simulations of materials by providing quantum-mechanical accuracy but orders-of-magnitude faster,
illustrated by ACM Gordon Bell prize (2020)~\cite{RN4} and finalist (2021)~\cite{RN5}.
State-of-the-art (SOTA) NNQMD model founded on group theory featuring rotational equivariance and local descriptors has provided much higher accuracy and speed than those models, thus named Allegro (meaning fast).
On massively parallel supercomputers, however,
it suffers a fidelity-scaling problem,
where growing number of unphysical predictions of interatomic forces prohibits simulations involving larger numbers of atoms for longer times.
Here, we solve this problem by combining the Allegro model with sharpness aware minimization (SAM) for enhancing the robustness of model through improved smoothness of the loss landscape.
The resulting Allegro-Legato (meaning fast and ``smooth") model was shown to elongate the time-to-failure $t_\textrm{\tiny failure}$,
without sacrificing computational speed or accuracy.
Specifically, Allegro-Legato exhibits much weaker dependence of time-to-failure on the problem size,
$t_{\textrm{\tiny failure}} \propto N^{-0.14}$
($N$ is the number of atoms) compared to the SOTA Allegro model $\left(t_{\textrm{\tiny failure}} \propto N^{-0.29}\right)$,
\textit{i.e.}, systematically delayed time-to-failure,
thus allowing much larger and longer NNQMD simulations without failure.
The model also exhibits excellent computational scalability and GPU acceleration on the Polaris supercomputer at Argonne Leadership Computing Facility.
Such scalable, accurate, fast and robust NNQMD models will likely find broad applications in NNQMD simulations on emerging exaflop/s computers,
with a specific example of accounting for nuclear quantum effects in the dynamics of ammonia to lay a foundation of the green ammonia technology for sustainability. \blfootnote{This work is published at International Supercomputing Conference 2023}
\end{abstract}

\section{Introduction}
Neural-network quantum molecular dynamics (NNQMD) simulations based on machine learning are revolutionizing atomistic modeling of materials by following the trajectories of all atoms with quantum-mechanical accuracy at a drastically reduced computational cost~\cite{RN1}.
NNQMD not only predicts accurate interatomic forces but also captures quantum properties such as electronic polarization~\cite{RN2} and electronic excitation~\cite{RN3},
thus the `$Q$' in NNQMD.
NNQMD represents one of the most scalable scientific applications on the current high-end supercomputers,
evidenced by ACM Gordon Bell prize winner in 2020~\cite{RN4} and finalist in 2021~\cite{RN5}.
A more recent breakthrough in NNQMD is drastically improved accuracy of force prediction~\cite{RN6} over those previous models,
which was achieved through rotationally equivariant neural networks based on a group theoretical formulation of tensor fields~\cite{RN7}.
The state-of-the-art (SOTA) accuracy has now been combined with a record speed based on spatially localized descriptors in the latest NNQMD model named Allegro (meaning fast)~\cite{RN8}.

\begin{figure}[ht]
  \centering
  \includegraphics[width=0.75\textwidth]{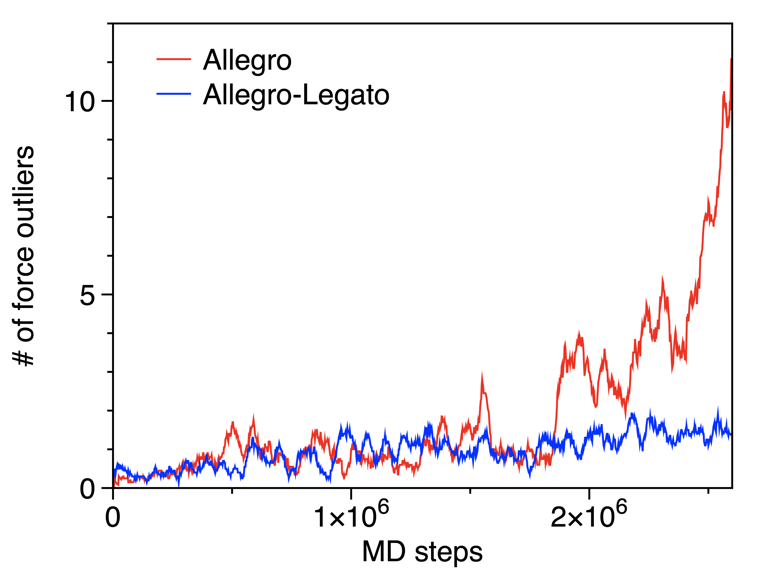}
  \caption{Number of outliers in atomic force inference during NNQMD simulation: As the simulation progresses,
    the dynamic of atoms becomes unstable due to an increasing number of unphysically large force values (over $5 \sigma$) predicted by the original Allegro model.
    This resulted in the eventual failure after $2.6 \times 10^6 \mathrm{MD}$ steps (red).
    On the other hand, the proposed model (Allegro-Legato) maintains a nearly constant number of outliers and the simulation stable (blue).}
  \label{fig:simulation_break}
\end{figure}

Despite its remarkable computational scalability, massively parallel NNQMD simulation faces a major unsolved issue known as fidelity scaling~\cite{RN9}.
In large-scale NNQMD simulations, small prediction errors can propagate and lead to unphysical atomic forces that degrade the accuracy of atomic trajectory over time.
These force outliers can even cause the simulation to terminate unexpectedly (Fig.~\ref{fig:simulation_break}).
As simulations become spatially larger and temporarily longer,
the number of unphysical force predictions is expected to scale proportionally,
which could severely limit the fidelity of NNQMD simulations on new exascale supercomputing platforms,
especially for the most exciting far-from-equilibrium applications~\cite{RN3,RN10}.

In this paper,
we solve the fidelity-scaling issue taking a cue from a recent development in machine learning.
Solving the fidelity-scaling issue requires robustness of the NNQMD model,
\textit{i.e.}, reduced number of unphysical force-prediction outliers when simulation trajectories encounter atomic configurations outside the training dataset.
It has been observed that the robustness of a neural-network model can be enhanced by sharpness-aware minimization (SAM)~\cite{RN11}
--- a training algorithm that regularizes the sharpness of the model
(\textit{i.e.}, the curvature of the loss surface) along with its training loss.
We thus apply SAM to train the fast Allegro model to smoothen its loss landscape,
thereby enhancing its robustness.
The resulting Allegro-Legato (meaning fast and ``smooth'') model is shown to increase the time-to-failure $t_{\textrm{\tiny failure}}$,
\textit{i.e.}, how many MD steps a NNQMD simulation can run under microcanonical ensemble,
while maintaining the same inference speed and nearly equal accuracy.
Specifically, Allegro-Legato exhibits much weaker dependence of time-to-failure on the problem size, $t_{\textrm{\tiny failure}} \propto N^{-0.14}$
($N$ is the number of atoms) compared to the SOTA Allegro model $\left(t_{\textrm{\tiny failure}} \propto N^{-0.29}\right)$,
thus allowing much larger and longer NNQMD simulations without failure. Along with this main contribution,
we find that the fidelity-scalability of the NNQMD model correlates with sharpness of the model more than the number of parameters in the model.~\footnote{Code is available at \href{https://github.com/ibayashi-hikaru/allegro-legato}{github.com/ibayashi-hikaru/allegro-legato}}

The fast and robust Allegro-Legato model has been implemented in our scalable parallel NNQMD code named RXMD-NN.
We have achieved a weak-scaling parallel efficiency of 0.91 on 480 computing nodes,
each with an AMD EPYC central processing unit (CPU) and four NVIDIA A100 graphics processing units (GPUs),
of the Polaris supercomputer at Argonne Leadership Computing Facility (ALCF).
The code has also achieved a 7.6-fold single-node performance acceleration using four GPUs over single 32-core CPU of Polaris.

Allegro-Legato allows much larger spatio-temporal scale NNQMD simulations than are otherwise possible.
Unlike MD simulation with heat bath often used in ``effective'' long-time sampling of molecular configurations (\textit{e.g.}, for protein folding),
which disrupts dynamic trajectories,
Allegro-Legato enables ``true'' long-time Hamiltonian dynamics that can be directly compared with fine vibrational modes observed in high-resolution spectroscopic experiments.
Specifically, we can now satisfy the prohibitive computational demand of accounting for subtle nuclear quantum effects in the dynamics of ammonia based on path-integral molecular dynamics,
which is essential for resolving a mystery in a recent high-resolution neutron-scattering experimental observation at Oak Ridge National Laboratory.
Synergy between the most advanced neutron experiment and leadership-scale NNQMD simulation lays a foundation of the green ammonia-based fuel technology for achieving a sustainable society.

\section{Method Innovation}
This section first summarizes (1) NNQMD simulation method, along with the SOTA Allegro model,
and (2) SAM for robust neural-network model training. We then present the key method innovation of SAM-enhanced Allegro model,
Allegro-Legato, followed by its scalable parallel implementation.
\subsection{Summary of Neural-Network Quantum Molecular Dynamics}
Molecular dynamics (MD) simulation follows time evolution of the positions $\left\{\mathbf{r}_i \mid i=1, \ldots, N\right\}$ (\textit{i.e.}, trajectories) of $N$ atoms,
\begin{equation}
  m_i \frac{d^2}{d t^2} \mathbf{r}_i=\mathbf{f}_i=-\frac{\partial}{\partial \mathbf{r}_i} E\left(\left\{\mathbf{r}_i\right\}\right)
\end{equation}
where $m_i$ and $\mathbf{f}_i$ are the mass of the $i$-th atoms and the force acting on it,
whereas $E$ is the interatomic potential energy that is dictated by quantum mechanics (QM).
In NNQMD, neural networks are trained to reproduce ground-truth $\mathrm{QM}$ values, $E\left(\left\{\mathbf{r}_i\right\}_t\right)$,
for a set of atomic configurations
$\left\{\left\{\mathbf{r}_i\right\}_t \mid t=1, \ldots, N_{\textrm{\tiny training}}\right\}$
($N_{\textrm{\tiny training}}$ is the number of training configurations)~\cite{RN1,RN2,RN3,RN4,RN5}.
In the SOTA Allegro model, the energy $E$ is composed of pairwise embedding energies, $E_{i j}$, between atomic pairs $(i, j)$ within a finite cutoff distance to preserve data locality~\cite{RN8}.
Key to the high accuracy of Allegro is that all energy terms are group-theoretically equivariant with respect to rotation, inversion and translation, \textit{i.e.}, to the Euclidean group $E(3)$~\cite{RN6,RN7}.
This is achieved by representing the energy in terms of tensors up to rank $\ell$ and tensor products using their irreducible representations.
In short, Allegro attains accuracy through group-theoretical equivariance and computational speed through data locality.

\subsection{Summary of Sharpness-Aware Minimization}
\label{sec:summary_sam}
Neural networks are trained by minimizing the loss function $L(\mathbf{w})$ where $\mathbf{w}$ represents the weight parameters of the neural network.
Design choice of optimization methods plays a crucial role in machine learning,
as it impacts various factors such as convergence speed and generalization performance~\cite{RN12}.
In particular, vulnerability to adversarial attacks is a problem unique to neural networks~\cite{RN13},
which has actively been studied in various fields such as computer vision~\cite{RN14} and natural language processing~\cite{RN4}.
Recent studies suggest that the fidelity-scalability in NNQMD can also be viewed as a robustness against ``adversarial attacks'' during large-scale simulations~\cite{RN15,RN16},
where atomic trajectories are ``attacked'' by the accumulated unphysical predictions, \textit{i.e.},
``adversarial perturbations'' throughout the long and large-scale simulation. Therefore, it is natural to expect that optimization methods for adversarial attack would enhance the fidelity-scalability in NNQMD.
Sharpness-aware minimization (SAM) is one of such robustness-enhancing methods proposed in the computer vision area~\cite{RN11}.
The key component of SAM is that it minimizes ``sharpness'' of the model defined as
\begin{equation}
  \label{eq:sharpness}
  \max_{\|\mathbf{\epsilon}\|_2 \leq \rho} \{L(\mathbf{w}+\mathbf{\epsilon})-L(\mathbf{w})\},
\end{equation}
where $\rho$ (the size of neighborhood) is a hyperparameter to define sharpness.
While computing the sharpness directly is infeasible, it has been shown that minimizing $L(\mathbf{w})+\max_{\|\mathbf{\epsilon}\|_2 \leq \rho}\{L(\mathbf{w}+\mathbf{\epsilon})-L(\mathbf{w})\}$
(training loss + sharpness) can be achieved through the following update rule:
\begin{equation}
  \mathbf{w}=\mathbf{w}-\eta \nabla_{\mathbf{w}^{\prime}} L\left(\mathbf{w}^{\prime}\right) \mid_{\mathbf{w}^{\prime}=\mathbf{w}+\rho \frac{\nabla_\mathbf{w} L(\mathbf{w})}{\left\|\nabla_\mathbf{w} L(\mathbf{w})\right\|}}\quad (\eta : \textrm{learning rate}),
\end{equation}
which utilizes first-order derivatives,
\textit{i.e.}, $\nabla_{\mathbf{w}} L(\mathbf{w})$.
This allows for the optimization of sharpness without the need for computationally expensive second-order derivatives.
\subsection{Key Innovation: Allegro-Legato: SAM-Enhanced Allegro}
As explained above, our hypothesis is that smoothened loss landscape through SAM enhances fidelity scaling of NNQMD.
To quantitatively test this hypothesis, we incorporate SAM into the training of the Allegro NNQMD model~\cite{RN8},
which entails SOTA accuracy and computational speed. We call the resulting SAM-enhanced Allegro model as Allegro-Legato.~\footnote{In music, Legato means smooth without sudden breaking between notes.}

To find an appropriate strength of sharpness regularization, SAM's hyperparameter $\rho$ is tuned so as to provide the most robust model,
from which we found that $\rho= 0.005$ gives the longest $t_{\textrm{\tiny failure}}$ in our setup.
For the small-scale simulation test, we used the LAMMPS, which is a widely used open-source MD simulation software (\href{https://www.lammps.org}{lammps.org}).
See Section~\ref{sec:training_details} for the detailed training settings.
\begin{table}[ht]
  \centering
  \begin{tabular}{l|ccccccc}
    \toprule
    $\rho$                       & $0.0$  & $0.001$ & $0.0025$ & $\mathbf{0.005}$ & $0.01$ & $0.025$ & $0.05$ \\
    \midrule
    $t_{\textrm{\tiny failure}}$ & $4020$ & $4030$  & $6420$   & $\mathbf{8480}$  & $4760$ & $4210$  & $3780$ \\
    \bottomrule
  \end{tabular}
  \vspace{0.5cm}
  \caption{\textbf{SAM strength $\rho$ \textit{vs.} time-to-failure} $t_{\textrm{\tiny failure}}$\textbf{:}
    We tune $\rho$ by conducting a grid search in the range of 0.001 to 0.05.
    A model with $\rho=0.005$ gives the largest $t_{\textrm{\tiny failure}}$ with a small-scale simulation $(N=432)$.}
  \label{table:table1}
\end{table}

\subsection{RXMD-NN: Scalable Parallel Implementation of Allegro-Legato NNQMD}
For the large-scale testing of computational and fidelity scaling,
we implement the proposed Allegro-Legato NNQMD model in our RXMD-NN software~\cite{RN3,RN9},
which is an extension of our scalable parallel reactive MD software, RXMD~\cite{RN17}.
RXMD-NN employs a hierarchical divide-and-conquer scheme to realize ``globally-scalable and local-fast'' (or ``globally-sparse and locally-dense'') parallelization \cite{RN18}:
(1) globally scalable spatial decomposition that is best suited for massively parallel computing platforms; and
(2) locally efficient linked-list decomposition and subsequent neighborlist construction to achieve the $O(N)$ computational complexity.
Interprocess communication is implemented using non-blocking application programming
interfaces (APIs) of Message Passing Interface (MPI) library, and the communication pattern is designed to be lock-free with minimal internode-data exchange.
While it is one of the most widely adapted strategies in large-scale MD applications,
this is particularly suitable for NNQMD algorithm to take advantage of the modern high-performance computing (HPC) architecture,
in which a few very powerful GPU cards do the heavy lifting by accelerating computationally demanding kernels while random memory access and out-of-order data processing are concurrently executed by many-core CPUs.
In RXMD-NN, CPU is responsible for the adjacency-list construction in parallel.
The constructed adjacency list, together with atom position and type information, is converted to PyTorch tensor object for force inference on GPUs.
RXMD-NN allows to control the computational granularity, such as the number of atoms per domain and domains per node, to find an ideal balance between horizontal and vertical scalability to utilize available hardware resources.

PyTorch has become a standard Python library in machine learning community due to its APIs for complex model architectures that enables highly efficient training and inference on GPU. However, production platforms such as HPC clusters, mobile devices, and edge nodes often demand a set of requirements that Python is not designed for, \textit{e.g.}, multithreading, low latency computing, and massively parallel distributed architectures. GPU Offloading of Allegro model is realized by TorchScript, which is statically typed intermediate representation to create serialized and optimizable $\mathrm{ML}$ models. The serialized model can be loaded from other programming language such as C++ allowing to be deployed in environments that are difficult for python codes to run without sacrificing multithreading and optimization opportunities.

\section{Results}
We test both fidelity and computational scalability of the proposed Allegro-Legato NNQMD model as implemented in the RXMD-NN code on a leadership-scale computing platform,
Polaris, at Argonne Leadership Computing Facility (ALCF).
\subsection{Experimental Platform}
We conduct numerical experiments on the Polaris supercomputer at ALCF.
It is a Hewlett Packard Enterprise (HPE) Apollo 6500 Gen 10+ based system consisting of two computing nodes per chassis,
seven chassis per rack, and 40 racks, with a total of 560 nodes. Each Polaris node has one $2.8 \mathrm{GHz}$ AMD EPYC Milan 7543P 32-core CPU with 512 GB of DDR4 RAM, four NVIDIA A100 GPUs with 40GB HBM2 memory per GPU,
two 1.6 TB of SSDs in RAID0 and two Slingshot network endpoints. Polaris uses the NVIDIA A 100 HGX platform to connect all 4 GPUs via NVLink, with a GPU interconnect bandwidth of $600 \mathrm{~GB} / \mathrm{s}$. Designed by Cray,
the Slingshot interconnect is based on high radix 64-port switches arranged in dragonfly topology, offering adaptive routing, congestion control and bandwidth guarantees by assigning traffic classes to applications. Polaris is rated at a production peak performance of 44 petaflops with node-wise performance at 78 teraflops for double precision.
\subsection{Fidelity-Scaling Results}
For the fidelity-scaling test, we trained Allegro and Allegro-Legato with $\ell=1$ and examined their robustness in terms of $t_{\textrm{\tiny failure}}$,
\textit{i.e.} the greater $t_{\textrm{\tiny failure}}$, the more robust.
The parameters of MD simulation for the test are carefully chosen so that each MD simulation is expected to fail within a reasonable time but not immediately.
While the constant-temperature ensemble method based on Nose-Hoover thermostat (\textit{i.e.}, NVT ensemble) is used to study thermal-equilibrium properties,
it could suppress and hidden unphysical model predictions by connecting atoms with an external thermostat. Microcanonical ensemble (NVE) method is the most rigorous test on the model robustness by simply integrating the equations of motion without an external control
(also it has broader applicability to non-equilibrium processes).
In each simulation instance, the liquid ammonia system is first thermalized at a temperature of $200 \mathrm{~K}$ using NVT ensemble for 1,000 steps.
We subsequently switch the ensemble to NVE and continue the simulation until it fails to determine $t_{\textrm{\tiny failure}}$.
The time step $\Delta t$ of 2 femto-seconds (fs) is chosen throughout the robustness test.
For each system size, over ten independent simulation instances are averaged to measure $t_{\textrm{\tiny failure}}$.

\begin{figure}[ht]
  \centering
  \includegraphics[width=0.75\textwidth]{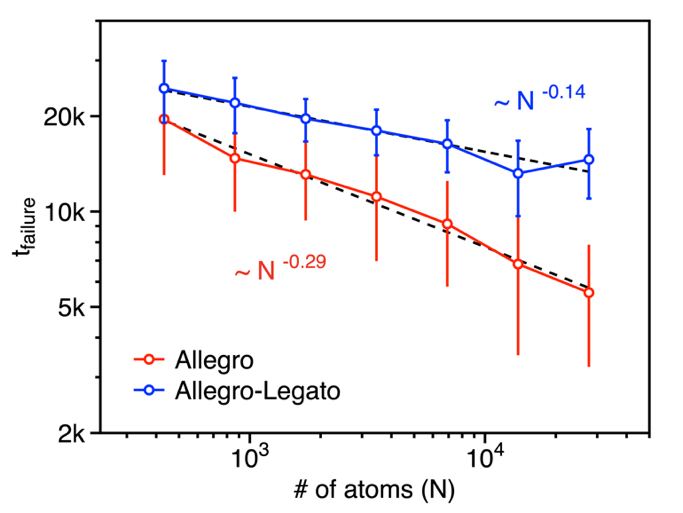}
  \caption{\textbf{Fidelity scaling of NNQMD simulation:} Here, $t_{\textrm{\tiny failure}}$ is measured using NVE ensemble with a timestep of $2 \mathrm{fs}$.
    Statistically improved $t_{\textrm{\tiny failure}}$ is observed in even the smallest system size,
    which is further pronounced as the system size increases.
    The exponent of power law fitting shows nearly a factor of two reduction using Allegro-Legato model.}
  \label{fig:fidelityscaling}
\end{figure}

To quantify fidelity scaling, we define a fidelity-scaling exponent $N^{-\beta}$ through the scaling relation,
\begin{equation}
  t_{\textrm{\tiny failure}}=\alpha N^{-\beta},
\end{equation}

where $\alpha$ is a prefactor.
A smaller $\beta$ value (\textit{i.e.}, weaker fidelity scaling) indicates delayed time-to-failure,
thus a capability to study larger spatiotemporal-scale processes accurately on massively parallel computers.
The Allegro-Legato model has drastically improved fidelity scaling, $\beta_{\textrm{\tiny Allegro-Legato}}=0.14<\beta_{\textrm{\tiny Allegro}}=0.29$ beyond statistical uncertainty
(see the error bars in Fig.~\ref{fig:fidelityscaling}),
thus systematically delaying time-to-failure.
\subsection{Computational-Scaling Results}
\begin{figure}[ht]
  \centering
  \includegraphics[width=0.75\textwidth]{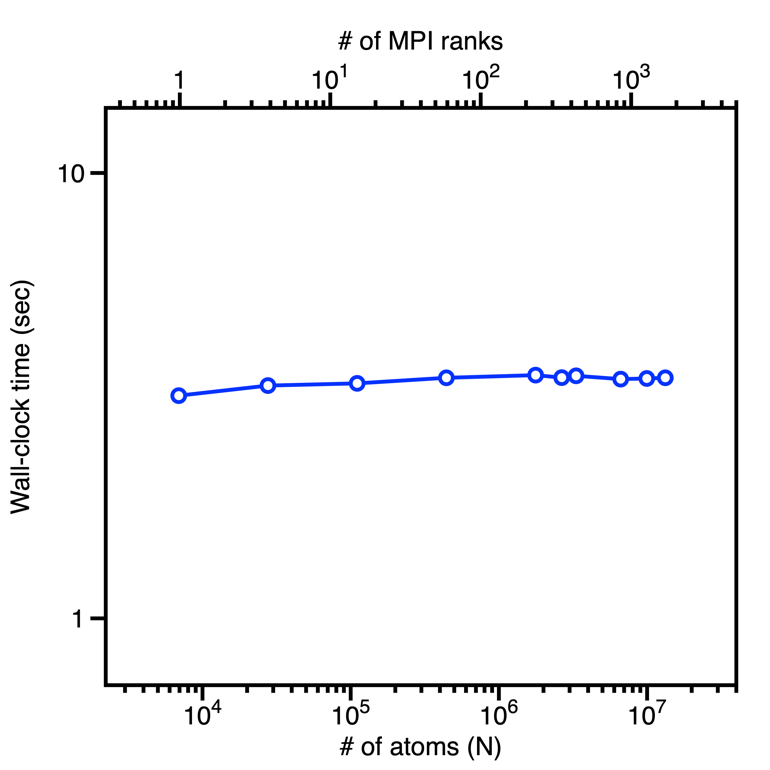}
  \caption{Wall-clock time of the RXMD-NN code per MD step, with scaled workloads 6,912 $P$ atom ammonia liquid using $P$ A100 GPUs $(P=$1, $\ldots$, 1,920).}
  \label{fig:scaling}
\end{figure}

We measure the wall-clock time per MD step with scaled workload, 6,912 $P$-atom ammonia system on $P$ MD domains.
In this test, each MD domain consists of 6,912 atoms that are offloaded to single GPU.
In addition to the force inference, the execution time includes the adjacency list construction, data transfer between host and GPU memory, and internode communication via network fabric.
Fig.~\ref{fig:scaling} shows wall-clock time as a function of $P$.
By scaling the problem size linearly with the number of GPUs, the runtime increases only slightly, indicating an excellent scalability.

Here, we quantify the parallel efficiency by defining the speed of $\mathrm{NNQMD}$ algorithm as the product of the total number of atoms multiplied by the number of MD steps executed per second.
The isogranular speedup is given by the speed on $P \mathrm{MD}$ domains relative to the speed of single domain as baseline.
The parallel efficiency of weak scalability thus is obtained by the isogranular speedup divided by $P$.
With the granularity of 6,912 atoms per domain, we have obtained an excellent weak-scaling efficiency, 0.91 for up to 13,271,040 atoms on 1,920 A100 GPUs.
Despite the relatively large granularity of 6,912 atoms per domain, we obtained a fast time-to-solution of 3.46 seconds per MD step enabling $25,000 \mathrm{MD}$ steps per day for production runs.

\begin{figure}[ht]
  \centering
  \includegraphics[width=0.75\textwidth]{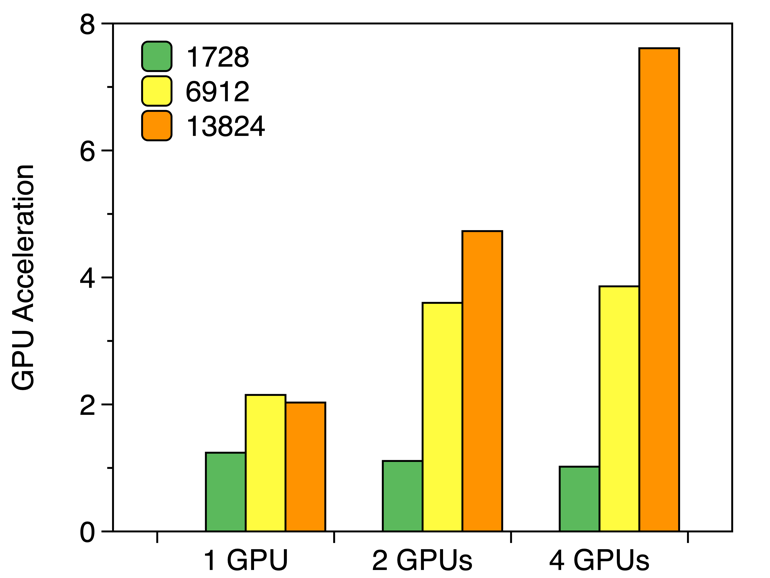}
  \caption{\textbf{GPU acceleration of NNQMD algorithm:} Three system sizes of $N=1728,6912$ and 13,824 atoms are examined. The histogram presents the reduction in wall-clock time per MD step over the runtime with $32 \mathrm{CPU}$ cores without GPU as reference.
    Detail of the benchmark platform as well as the GPU and CPU architectures are presented in the main text. We have achieved $7.6 \mathrm{x}$ speedup using four GPUs with $N=$13,824 atoms.}
  \label{fig:gpu_scaling}
\end{figure}

Fig.~\ref{fig:gpu_scaling} shows GPU acceleration of NNQMD algorithm on single Polaris node. The histogram presents the reduction in wall-clock time per MD step (averaged over $10 \mathrm{MD}$ steps) using the runtime obtained with CPU only (32 cores with 32 threads) as baseline.
Here, we examined: (1) three system sizes of $N=$1,728,6,912, and 13,824 ammonia atoms; and (2) three domain decomposition such as single, double and quadruple subdomains.
Atoms in each domain are assigned to one GPU. With $N=$1,728 system,
we observe a marginal GPU acceleration up to 1.24x speedup,
which has been substantially improved with greater system sizes. We have achieved a 7.6x speedup using $N=$13,824 atom system with four subdomains.

\section{Discussion}
While SAM-enhanced Allegro model, Allegro-Legato, has achieved improved robustness over the SOTA Allegro model as shown in the previous section, we here discuss the imprecation of SAM training to other aspects such as accuracy and computational speed.
\subsection{Simulation Time}
First, MD simulation time is not affected by SAM since SAM only applies to the training stage but not the inference stage in MD simulation. Table~\ref{table:table2} compares the simulation time per MD time step for the baseline Allegro model and the proposed Allegro-Legato model. Hereafter, we use the default value, $\ell=1$, for the maximum tensor rank, thus the same number of parameters for the two models. The simulation time is identical for both models within the measurement uncertainty due to nondedicated access to the experimental platform.

As a comparison, Table~\ref{table:table2} also shows the baseline Allegro model with two other tensor ranks, $\ell=0$ and 2. Larger $\ell$ generates more accurate but larger models (\textit{i.e.}, larger numbers of parameters) and hence incur longer simulation times. Based on the accuracy/computational-cost trade-off, production NNQMD simulations with the Allegro model typically use $\ell=1$.
\begin{table}[ht]
  \centering
  \begin{tabular}{l|c|c}
    \toprule
    Model                                  & \# of parameters & Time/step $(\mathrm{ms})$ \\
    \midrule
    Allegro                                & 133,544          & 916                       \\
    Allegro-Legato                         & 133,544          & 898                       \\
    \midrule
    \midrule
    \multicolumn{2}{l}{ Reference Models } &                                              \\
    \midrule
    Allegro $(\ell=0)$                     & 95,656           & 395                       \\
    Allegro $(\ell=2)$                     & 183,720          & 2,580                     \\
    \bottomrule
  \end{tabular}
  \vspace{0.5cm}
  \caption{\textbf{Simulation-time comparison:} As SAM only applies to the training stage and does not modify the size of architecture, the computational cost for simulation is not affected.}
  \label{table:table2}
\end{table}
\subsection{Training Time}
As mentioned in Section~\ref{sec:summary_sam}, SAM's shortcoming is that it requires more computation time than the base optimizer,
because each epoch has to compute the first-order gradients twice. However, in our setting,
SAM converges faster than the default optimizer, and thus the total training time is not significantly affected (Table~\ref{table:table3}).
As references, we also measured the training time of Allegro models with different maximum tensor ranks, $\ell=0$ and 2 and we observed that the training cost increases drastically for larger $\ell$.
In summary, Allegro-Legato improves the robustness of Allegro without incurring extra training cost.
\begin{table}[ht]
  \centering
  \begin{tabular}{l|c|c|c}
    \toprule
    Model                                  & Total time (hours) & Per-epoch time (seconds) & Epochs \\
    \midrule
    Allegro                                & 11.1               & 248                      & 161    \\
    Allegro-Legato                         & 13.6               & 433                      & 113    \\
    \midrule
    \midrule
    \multicolumn{3}{l}{ Reference Models } &                                                        \\
    \midrule
    Allegro $(\ell=0)$                     & 4.4                & 127                      & 127    \\
    Allegro $(\ell=2)$                     & 19.6               & 636                      & 111    \\
    \bottomrule
  \end{tabular}
  \vspace{0.5cm}
  \caption{\textbf{Training-time comparison:} Although SAM takes longer per-epoch training time, it converges faster and thus does not significantly affect total training time.
    Compared to the reference training times of variations of Allegro models, the extra training cost of Allegro-Legato is negligible.}
  \label{table:table3}
\end{table}
\subsection{Model Accuracy}
While faithful reproduction of system energy is necessary to properly guide model training,
the most crucial to MD simulations is accurate force prediction.
We obtained the validation error in atomic force as 15.9 (root mean square error, RMSE) and 11.6 (mean absolute error, MAE) with Allegro-Legato $(\ell=1)$ model, and 14.7 (RMSE) and 10.7 (MAE) with the original Allegro model $(\ell=1)$, respectively.
All error values are in a unit of $\mathrm{meV} / \AA$. Chmiela et al. recently provided a guideline that MAE required for reliable $\mathrm{MD}$ simulations is $1 \mathrm{kcal} / \mathrm{mol} / \AA$, which corresponds to $43.4 \mathrm{meV} / \AA$ ~\cite{RN19}.
Although Allegro-Legato incurs a slight increase in the force prediction error (about $8 \%$ in the liquid ammonia dataset) compared to the original Allegro model,
the obtained force error is about a factor four smaller than the guideline for reliably performing MD simulations. Namely, Allegro-Legato improves the robustness without sacrificing accuracy.

\subsection{Implicit Sharpness Regularization in Allegro}
While we propose to explicitly control the sharpness of models, we found that one control parameter in the baseline Allegro model (\textit{i.e.}, maximum rank of tensors to represent features) implicitly regulate the sharpness of the model.
In Table~\ref{table:table4}, besides our Allegro-Legato model having smaller sharpness,
Allegro $\ell=1,2$ models have significantly smaller sharpness and higher $t_{\textrm{\tiny failure}}$ compared to Allegro $\ell=0$ model.
Namely, Allegro with higher $\ell$ implicitly regularizes sharpness, resulting in higher robustness (\textit{i.e.}, larger $t_{\textrm{\tiny failure}}$),
but with increasing computational cost. Allegro-Legato $(\ell=1)$ model achieves the same level of sharpness as Allegro $(\ell=2)$ model with much less computing time;
see Tables~\ref{table:table2} and \ref{table:table3}.
\begin{table}[ht]
  \centering
  \begin{tabular}{c|cccc}
    \toprule
    \multirow{2}{*}{Model} & \multirow{2}{*}{Allegro $(\ell=0)$} & \multirow{2}{*}{Allegro $(\ell=1)$} & \multirow{2}{*}{Allegro $(\ell=2)$} & Allegro-Legato       \\
                           &                                     &                                     &                                     & $(\ell=1)$           \\
    \midrule
    Sharpness              & $5.0 \times 10^{-4}$                & $3.2 \times 10^{-4}$                & $9.8 \times 10^{-5}$                & $1.2 \times 10^{-4}$ \\
    \bottomrule
  \end{tabular}
  \vspace{0.5cm}
  \caption{\textbf{Implicit sharpness regularization by Allegro:}
    While our Allegro-Legato model has smaller sharpness than Allegro,
    Allegro models with larger $\ell$ have progressively smaller sharpness.
    Here, we measure sharpness, $\max _{\|\mathbf{\epsilon}\|_2 \leq \rho}\{L(\mathbf{w}+\mathbf{\epsilon})-L(\mathbf{w})\}$,
    by taking maximum of 1,000 independent random samples around the 0.05-neighborhood of each minimum.}
  \vspace{-1.0cm}
  \label{table:table4}
\end{table}

\begin{figure}[ht]
  \centering
  \includegraphics[width=0.75\textwidth]{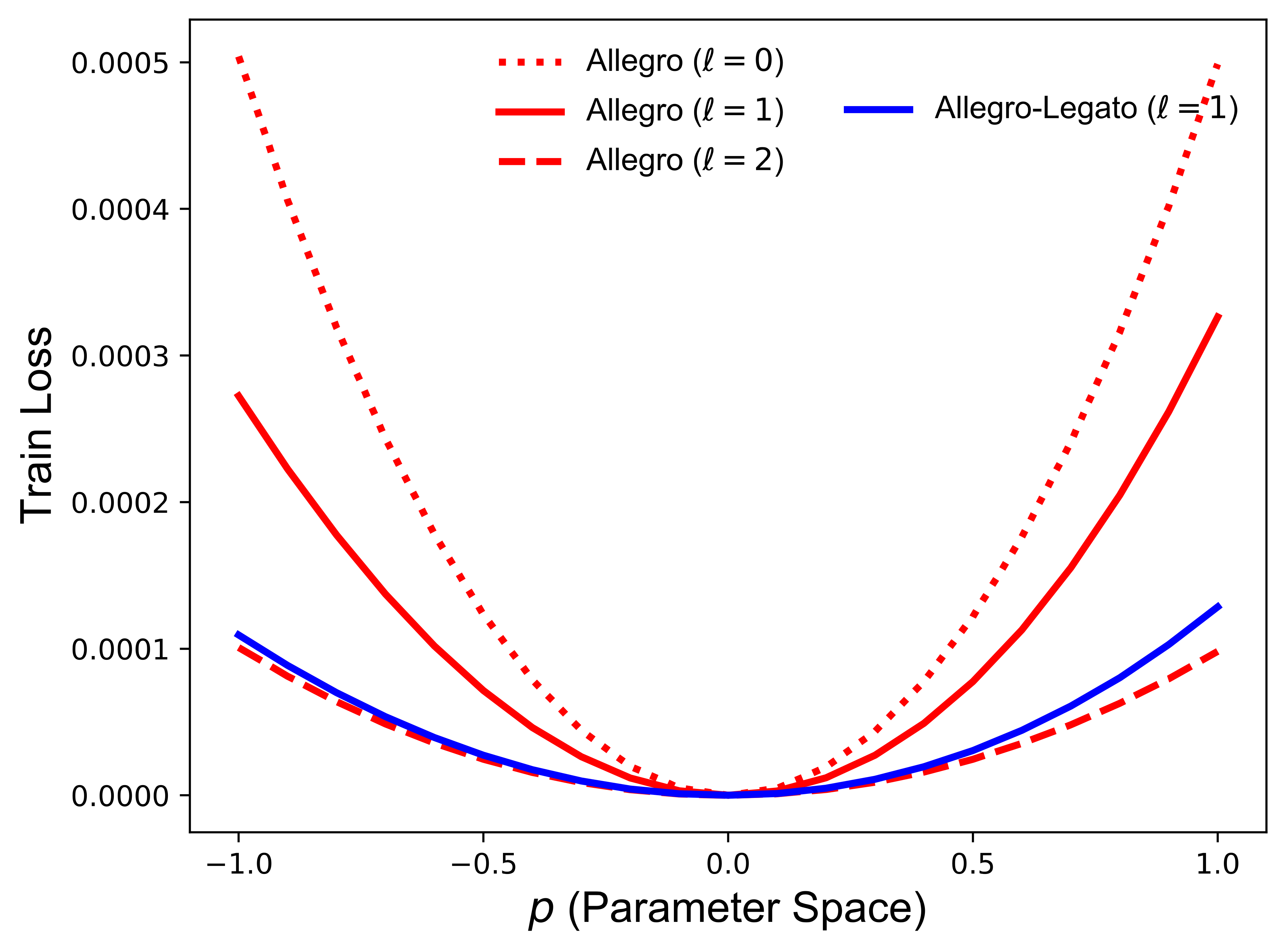}
  \caption{\textbf{Loss surface visualization:} One dimensional visualization of loss surface of each model.
    Following the definition of sharpness (Eq.~\ref{eq:sharpness}), we randomly sample a vector, $\mathbf{d}$,
    that gives the sharpness direction to compute $L(\mathbf{w}+p \mathbf{d})$ for $p \in[-1,1]$.}
  \label{fig:loss_surface}
\end{figure}

Fig.~\ref{fig:loss_surface} visualizes the loss surface of Allegro ($\ell$=0,1, and 2) and Allegro-Legato $(\ell$ $=1$) models.
The figure confirms: (1) progressive smoothening (\textit{i.e.}, smaller sharpness) for larger $\ell$ within the Allegro model due to implicit regularization through accuracy but with increasing computational cost;
and (2) explicit smoothening of Allegro-Legato through SAM over Allegro with the same $\ell$ without extra computational cost.
\subsection{Training Details}
\label{sec:training_details}
Lastly, we provide detailed training configuration for completeness (Table~\ref{table:table5}).
For fair comparison, we used the default hyperparameters that are released as the SOTA model and SAM training uses the default optimizer as its base optimizer.
\begin{table}[ht]
  \centering
  \begin{tabular}{l|c}
    \toprule Material type                                            & Liquid $\textrm{NH}_3$                                              \\
    \midrule Number of atoms per a configuration                      & 432                                                                 \\
    \midrule \# of training examples ($N_{\textrm{\tiny training}}$)  & 4,500                                                               \\
    \midrule \# of validation examples                                & 500                                                                 \\
    \midrule $\mathbf{r}_{\textrm{\tiny max}}$ for cutoff             & 6.0                                                                 \\
    \midrule Maximum tensor rank $(\mathbf{\ell})$                    & 1                                                                   \\
    \midrule Batch size                                               & 4                                                                   \\
    \midrule Peak learning rate                                       & $2 \mathrm{e}-3$                                                    \\
    \midrule Learning rate decay                                      & ReduceLROnPlateau                                                   \\
    \midrule Learning rate scheduler patience                         & 50                                                                  \\
    \midrule Learning rate scheduler factor                           & 0.5                                                                 \\
    \midrule (Base) Optimizer                                         & Adam                                                                \\
    \midrule Adam's ($\mathbf{\beta}_1, \mathbf{\beta}_{\mathbf{2}}$) & $(0.9,0.999)$                                                       \\
    \midrule  Loss function                                           & Per atom MSE                                                        \\
    \midrule  Loss coefficient (force, total energy)                  & $(1.0, 1.0)$                                                        \\
    \midrule   Stopping criterion                                     & $\Delta L_{\textrm{validation}} \leq 3 \mathrm{e}-3$ for 100 epochs \\
    \bottomrule
  \end{tabular}
  \vspace{0.5cm}
  \caption{
    \textbf{Detailed training setting:} All training setups in this paper adopt these parameters unless otherwise noted.
  }
  \label{table:table5}
\end{table}
\section{Applications}
The improved robustness of the proposed Allegro-Legato model,
while preserving the SOTA accuracy and computational speed of Allegro, enables large spatio-temporal scale NNQMD simulations on leadership-scale computers.
A compelling example is the study of vibrational properties of ammonia.
Development of dynamical models that accurately reproduce the vibrational spectra of molecular crystals and liquids is vital for predictions of their thermodynamic behavior,
which is critical for their applications in energy, biological, and pharmaceutical systems~\cite{RN20}.
In particular, there has been growing development of green ammonia-based fuel technologies for sustainable society over the past few years.
Ammonia $\left(\mathrm{NH}_3\right)$ has a higher energy density than even liquid hydrogen, but ammonia can be stored at a much less energy-intensive $-33{ }^{\circ} \mathrm{C}$ versus $-253^{\circ} \mathrm{C}$,
and thanks to a century of ammonia use in agriculture, a vast ammonia infrastructure already exists~\cite{RN21}. Over 180 million metric tons of ammonia is produced annually, and 120 ports are equipped with ammonia terminals~\cite{RN21}.
Development of technologies based on ammonia will be reliant on our ability to understand and model the complex physical and chemical interactions that give rise to its unique properties.

\begin{figure}[ht]
  \centering
  \includegraphics[width=0.95\textwidth]{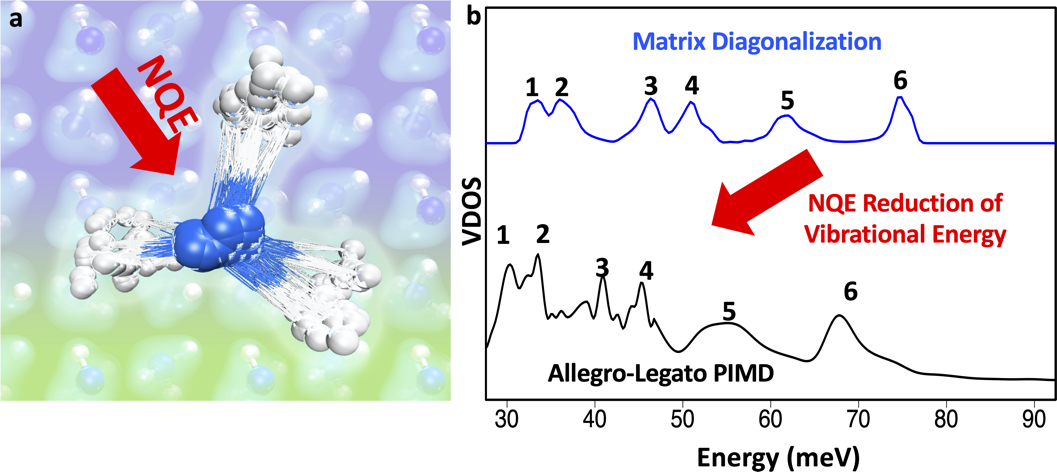}
  \caption{
    \textbf{Computed vibrational spectra of ammonia:} (a) While typical first-principles simulation treats atoms classically and electrons quantum-mechanically, PIMD simulation uses multiple replicas of each atom to mimic nuclear quantum effect (NQE). (b) Top curve shows vibrational spectrum computed at zero temperature without $\mathrm{NQE}$, while bottom at finite temperature with Allegro-Legato PIMD simulation. With the inclusion of NQE, Allegro-Legato PIMD correctly shows softening of high-energy inter-molecular modes expected at finite temperature and explains high-end neutron-scattering observations.
  }
  \label{fig:application}
\end{figure}

There are multiple complicating factors that require careful considerations such as nuclear quantum effects (NQEs)
and its coupling with vibrational anharmonicity when developing computational frameworks that accurately describe vibrational properties~\cite{RN20}.
Standard first-principles calculations for vibrational properties only treat electrons quantum mechanically and vibrational properties can be determined by Fourier transform and matrix diagonalization of the unit-cell Hessian, which is at most on the order of a few 100 entries~\cite{RN22}.
Evaluating the role of NQEs and its coupling with vibrational anharmonicity is done in the so-called path integral MD (PIMD) approach, which samples the quantum partition function for the entire quantum system~\cite{RN23,RN24}.
This requires long-time simulations of a large number of replicas of large MD systems that are harmonically coupled to each other as interacting ring-polymers,
especially at low temperatures~\cite{RN23,RN24}.
The background of Fig.~\ref{fig:application}a shows a typical first principles-based simulation,
where the atoms are treated classically, and the electron charge density is treated quantum-mechanically to compute atomic forces, which is illustrated as blue iso-surfaces.
In the foreground we have highlighted one $\mathrm{NH}_3$ molecule from a PIMD simulation of the same atomic configuration,
where each atom has 32 replicas that are harmonically coupled together.
The computation of the replica simulations is embarrassingly parallel, with only fixed nearest replica communication,
and the major cost is computing the energy and forces for the atoms within each replica simulation,
which is typically done from first principles. However, our Allegro-Legato model with enhanced robustness allows for stable long-time $\mathrm{MD}$ simulations at near quantum accuracy,
and thus can replace expensive first-principles calculations in the PIMD simulations, which would make accurate evaluation of ammonia's low energy intermolecular vibrational modes intractable.

We have performed massively parallel PIMD simulations with our Allegro-Legato model,
computing the energy and forces within each replica simulation to evaluate the phonon spectra for inter-molecular modes of ammonia.
The Allegro-Legato model is found to produce the expected softening of high-energy modes at finite temperature with inclusion of nuclear quantum effects in comparison to standard matrix diagonalization within the harmonic approximation,
which is illustrated in Fig.~\ref{fig:application}b.
In particular, reduction of the energy of the vibrational modes in the $30-90 \mathrm{meV}$ is consistent with high-end neutron experiments for the vibrational spectrum performed by the authors at Oak Ridge National Laboratory in the last summer (these results will be published elsewhere).

\section{Related Work}
There has been an explosion in the development and application of NNQMD simulations~\cite{RN1,RN2,RN3,RN6,RN8} and their scalable parallel implementation~\cite{RN4,RN5}.
On the other hand, it was only recently that the robustness of NNQMD was quantified in terms of time-to-failure $t_{\textrm{\tiny failure}}$\cite{RN25} and its deteriorating reduction with the problem size (\textit{i.e.}, fidelity-scaling problem) was pointed out~\cite{RN9}.
This work is the first to: (1) formally quantify the fidelity scaling by introducing the fidelity-scaling exponent $\beta$ through $t_{\textrm{\tiny failure}} \propto N^{-\beta}$ ($N$ is the number of atoms);
and (2) propose the solution to the fidelity-scaling problem using sharpness-aware minimization.

Robustness against adversarial attacks is a central and widely studied issue in machine learning~\cite{RN4,RN13,RN14}.
Compared to typical adversarial attacks, it is nontrivial to generate adversarial perturbations for NNQMD.
This is because the attack we consider is not only focused on the accuracy of the model, but also on the time to failure ($t_{\textrm{\tiny failure}}$) of the model,
which can only be determined through long-time simulations~\cite{RN15,RN16}.
Generative adversarial network (GAN) is one possible approach for sampling molecular configurations in a learning-on-the-fly setting~\cite{RN26}.
However, we remark that the real strength of MD simulation is its ability to compute dynamic correlations that can directly explain high-resolution spectroscopic experiments,
which requires a long uninterrupted Hamiltonian trajectory,
to which adversarial networks are generally not applicable.
In this domain, Allegro-Legato thus provides a unique solution.

\section{Conclusion}
We have introduced the proposed SAM-based solution to the fidelity-scaling problem into the Allegro NNQMD model~\cite{RN8}
which represents the state-of-the-art accuracy and speed.
The resulting Allegro-Legato model has drastically improved fidelity scaling by exhibiting a significantly lower exponent, $\beta_{\textrm{\tiny Allegro-Legato}}$$=0.14$$<\beta_{\textrm{\tiny Allegro}}=0.29$,
thus systematically delaying time-to-failure. Such improved fidelity scaling is central to ensure that meaningful scientific knowledge is extracted from large-scale simulations on leadership-scale parallel computers.
Our scalable parallel implementation of Allegro-Legato with excellent computational scaling and GPU acceleration combines accuracy, speed, robustness and scalability,
thus allowing practical large spatiotemporal-scale NNQMD simulations for challenging applications on exascale computing platforms.
\section{Acknowledgments}
This work was supported as part of the Computational Materials Sciences Program funded by the U.S. Department of Energy, Office of Science, Basic Energy Sciences, under award number DE-SC0014607. H.I. and K.N. were partially supported by an NSF grant, OAC-2118061.
The simulations were performed at the Argonne Leadership Computing Facility under the DOE INCITE program, while scalable code development was supported by the Aurora ESP program. The authors acknowledge the Center for Advanced Research Computing at the University of Southern California for providing computing resources that have contributed to the research results reported within this publication.
We are grateful to Dr. Makiko Hirata for valuable discussions regarding Allegro-Legato.

\bibliographystyle{splncs04}
\bibliography{main}
\end{document}